\documentclass[prl,twocolumn,superscriptaddress]{revtex4-2}  
\usepackage{graphicx,amssymb}
\usepackage{color,ulem}
\usepackage{bm}   

\usepackage{bm}% bold math
\usepackage{amssymb,amsmath}
\usepackage{amsthm}

\usepackage{latexsym}
\usepackage{graphicx,amssymb}
\usepackage{color,ulem}
\usepackage{float}
\usepackage{siunitx}
\usepackage{mathrsfs} 
\usepackage{soul}
\usepackage{hyperref}
\usepackage{graphicx}% Include figure files
\usepackage{dcolumn}% Align table columns on decimal point
\usepackage{bm}% bold math
\usepackage[utf8]{inputenc}
\usepackage{amsfonts,amssymb,amsmath}
\usepackage[mathscr]{eucal}
\usepackage{epsfig}

\begin{document}
% \preprint{AIP/123-QED}

\title{Comparison of Lindblad and circuit approaches for quantum heat transport}
	\author{Bayan Karimi}
\address{Pritzker School of Molecular Engineering, University of Chicago, Chicago IL 60637, USA}
\affiliation{Pico group, QTF Centre of Excellence, Department of Applied Physics, Aalto University School of Science, P.O. Box 13500, 00076 Aalto, Finland}

\date{\today}

\begin{abstract}
We compare two popular models applicable to analyzing heat transport by thermal microwave photons in quantum circuits. The first model is derived from a weak-coupling Lindblad master equation, with transition rates determined by Fermi's golden rule induced by thermal dissipation sources. The second approach employs a circuit model, where thermal Johnson–Nyquist noise generated by dissipative elements introduces currents, and consequently Joule power, in other parts of the circuit. This leads to a Landauer type expression of heat transport where the transmission coefficient is proportional to the transconductance in the circuit. We find that the two models yield identical results in a linear circuit in the weak coupling limit with an analytic expression of power in an archetypal circuit of a cavity mediating heat between two baths. Our analysis yields a quantitative assessment of the range of validity of the weak coupling assumption in a circuit. Due to the correspondence of the two results, we feel confident in applying the weak coupling Lindblad model also for analyzing heat transport in quantum circuits consisting, e.g. of qubits and/or non-linear resonators. 
 
\end{abstract}

\maketitle
%{\sl Introduction:} 
Quantum heat transport by microwave photons in a linear circuit is normally treated by circuit models, where a Landauer type expression with (heat) current governed by the transmission coefficient is given by the electrical transconductance between the hot and cold elements~\cite{Schmidt,Meschke,Pascal,Dimas2023}. This method is, however, not directly applicable for nonlinear circuits, for instance those consisting of qubits or anharmonic oscillators. In such a situation one may resort to methods based on the density operator~\cite{Davies,Hartmann} that can be obtained, e.g. from a Lindblad type master equation with rates given by the Fermi's golden rule expressions~\cite{Breuer,Lindblad,Sudarshan,Pascazio,Trushechkin,Rudner,Horodecki,Schnell}. This approach is generally applicable for systems where the coupling between the bath and the quantum system is weak. In this paper, we investigate an archetypal two-bath system, where a cavity in the form of a $LC$ oscillator mediates the heat between the two. This system has been experimentally studied in the so-called quantum heat valve configuration~\cite{Ronzani,Partanen,Pekola,Xu,Thomas,Antti2026}. We demonstrate that in the limit of weak coupling of the cavity to the two baths, the two methods yield identical results, and the heat current is given by a universal expression, where the quantitative value of the heat conductance is determined by the circuit parameters directly. This result lends confidence towards analyzing also nonlinear circuits quantitatively using the quantum master equation approach, i.e. in the regime where the standard circuit model is not applicable. As a bonus, the results demonstrate the validity range of the weak coupling Lindblad model in terms of the physical coupling parameter.

\begin{figure}[h!]
	\centering
	\includegraphics [width=\columnwidth] {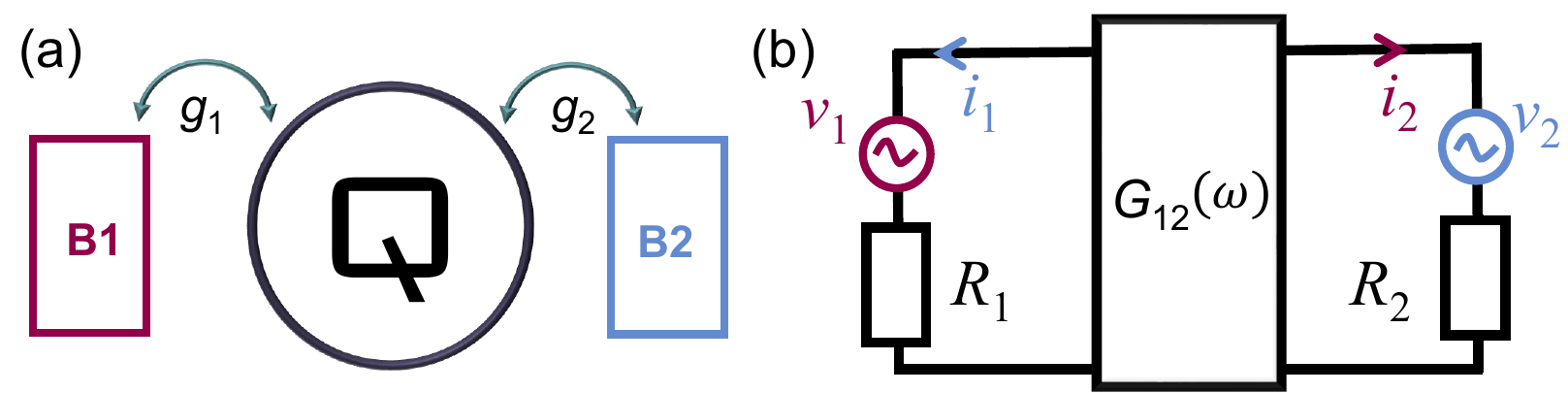}
	\caption{The two-bath setup in an electric circuit. In the Lindblad model in (a), heat between baths B1 and B2 is transported via emission and absorption of microwave photons by each of them on the quantum circuit ``Q" in between. In the circuit model in (b) each resistor bath generates thermal noise voltage that is converted into a noise current in the opposite resistor, thus transporting heat between the two.
		\label{fig1}}
\end{figure}

Figure \ref{fig1} presents the main idea of this work. There are two seemingly different methods to find the thermal conductance in a generic linear circuit: the ``quantum" approach based on weak coupling transition rates in Fig.~\ref{fig1}\,(a), which we call the Lindblad-model, and the ``microwave circuit" approach in Fig.~\ref{fig1}\,(b).  

In the Lindblad model, the quantum system is placed between the two baths. These baths induce transitions in the system, and the transition rates are given by golden rule expressions in the weak coupling picture arising from the equilibrium thermal noise of each resistor acting as the bath. An example will be given below. In general, the populations, i.e., the diagonal elements $\rho_n$ (for the energy eigenstate $n$ of the system) of the density operator evolve as
\begin{equation} \label{genmaster}
	\dot \rho_n =\sum_{i\neq n} \Gamma_{i\rightarrow n} \rho_i - \sum_{j\neq n} \Gamma_{n\rightarrow j} \rho_n.	
\end{equation}
Here, the total transition rate from state $r$ to state $s$ ($\Gamma_{r\rightarrow s}$) is the sum of the contributions from the two uncorrelated baths,
\begin{equation}\label{transitionrates1}
    \Gamma_{r\rightarrow s}
    = \Gamma^{(1)}_{r\rightarrow s}
    + \Gamma^{(2)}_{r\rightarrow s},
\end{equation}
where $\Gamma^{(i)}_{r\rightarrow s}$ denotes the transition rate induced by bath $i$ $(i=1,2)$. For obtaining steady state populations, we solve Eq. \eqref{genmaster} by setting $\dot \rho_n =0$. The power $P_{B}$ into each bath $B=1,\,2$ is given by
\begin{equation} \label{genpower}
	P_B=\sum_{k,l} \rho_k (E_k-E_l)\Gamma^{(B)}_{k \rightarrow l},	
\end{equation}
where $E_r$ is the energy of the $r$:th eigenstate.

\begin{figure}
		\centering
		\includegraphics [width=\columnwidth] {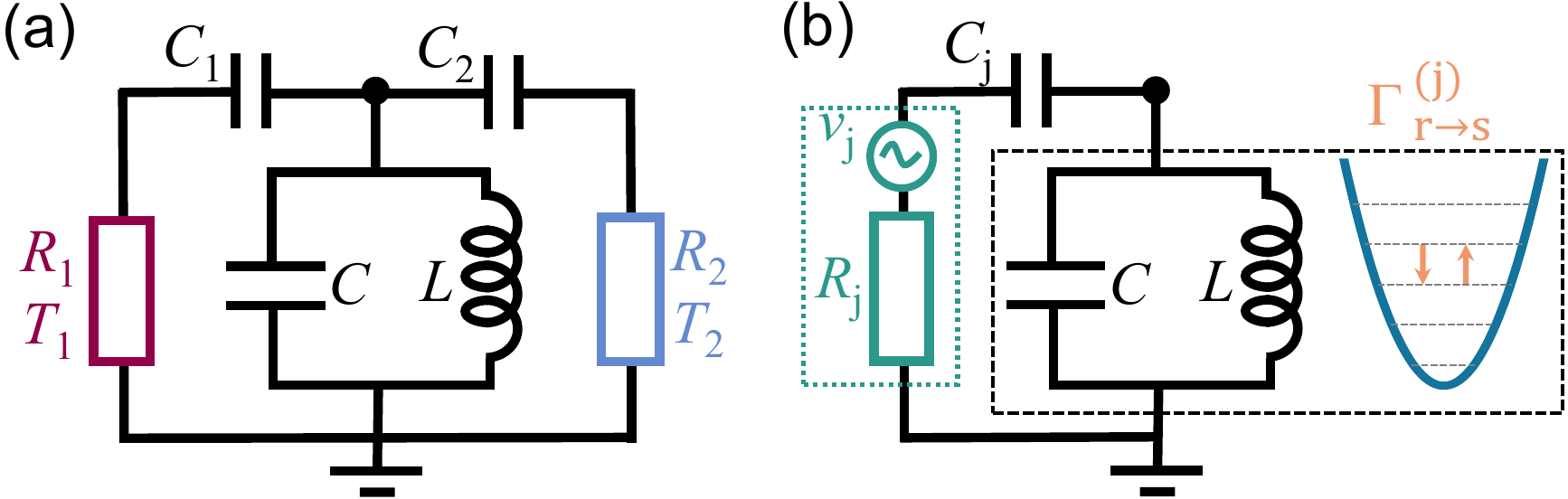}
		\caption{The exemplary circuit to assess the two models. (a) The quantum circuit between the baths, namely the two resistors $R_j$ maintained at temperatures $T_j$ for $j=1,\,2$, is represented by a harmonic oscillator ($LC$-circuit) coupled to the baths via the coupling capacitances $C_j,\,j=1,\,2$. (b) Illustration of how the resistor $R_j$ acts as a thermal noise source acting on the $LC$-circuit which forms a quantum harmonic oscillator. 
			\label{fig2}}
	\end{figure}

For the example circuit in Fig.~\ref{fig2}\,(a), following the uncorrelated noise idea, we consider each half of the circuit separately, as shown in Fig.~\ref{fig2}\,(b). Here, a resistor $R_j$ in series with coupling capacitance $C_j$ acts as a noise source for a parallel oscillator with inductance $L$ and capacitance $C$. By elementary analysis, we find the classical energy loss rate of the harmonic oscillator as 
$\dot E \approx -\frac{R_j}{Z_0}(\frac{C_j}{C_\Sigma})^2\omega_0 E$, where $Z_0=\sqrt{L/C_\Sigma}$, $C_\Sigma =C+C_j$ and $\omega_0=1/\sqrt{LC_\Sigma}$, valid in the weak-coupling limit $R_j \ll 1/(\omega_0 C_j)$. 
We next compare this to the quantum mechanical heat transfer rate to the resistor for the same system occupying the $n$th level with energy $E=n\hbar\omega_0$ (ignoring naturally the zero point energy). We have
$\dot E =-\Gamma_{n\rightarrow n-1}^{(j)} \hbar\omega_0$. The classical correspondence principle allows us to identify $\Gamma_{n\rightarrow n-1}^{(j)} = n \frac{R_j}{Z_0}(\frac{C_j}{C_\Sigma})^2 \omega_0$, and the coupling constant is
\begin{equation} \label{coupling} 
	g_j=\frac{R_j}{Z_0}(\frac{C_j}{C_\Sigma})^2 
\end{equation} 
yielding 
\begin{equation} \label{transition-rates}
	\Gamma_{n\rightarrow n-1}^{(j)} = n g_j \frac{\omega_0}{1-e^{-\beta_j\hbar\omega_0}}, \,\,\, \Gamma_{n-1\rightarrow n}^{(j)} = e^{-\beta_j\hbar\omega_0}	\Gamma_{n\rightarrow n-1}^{(j)},
\end{equation} 
for standard relaxation and excitation rates of a harmonic oscillator coupled to a resistor at inverse temperature $\beta_j=(k_B T_j)^{-1}$. 

We now explicitly include the two uncorrelated noise sources separately. Following the proper indexing shown in Fig. \ref{fig2}\,(a), we may write the master equation for the diagonal element $\rho_n$  of the density matrix of the harmonic oscillator of the $n$th energy level as 	
\begin{eqnarray}\label{densityM-n}
		&&\dot{\rho}_n=\Gamma_{n-1\rightarrow n}\rho_{n-1}+\Gamma_{n+1\rightarrow n}\rho_{n+1}\nonumber\\&&~~~~~~~-(\Gamma_{n\rightarrow n-1}+\Gamma_{n\rightarrow n+1})\rho_n.
	\end{eqnarray}
We find steady state populations $(\dot{\rho}_n=0)$, and the power to bath $j$ with constant temperature difference as 
\begin{equation} \label{power}
P_1=-P_2=\hbar\omega_0\sum_{n=0}^{\infty}\rho_n(\Gamma_{n\rightarrow n-1}^{(1)}-\Gamma_{n\rightarrow n+1}^{(1)}).
\end{equation}
Although this procedure allows to find, at least numerically, the heat currents for any temperatures, we limit ourselves here to linear response to find the thermal conductance. We set $\beta_1 \equiv \beta \equiv 1/(k_B T)$ and $\beta_2=\beta +\delta \beta$, where $\delta \beta =-1/(k_B T^2) \delta T$ for a small temperature difference $\delta T$. Without further approximations we then obtain the population of the ground state as 
\begin{equation}\label{rho-ground}
	{\rho_0}=1-(1-\frac{g_2}{g_1+g_2}\delta\beta\hbar\omega_0)e^{-\beta\hbar\omega_0},
\end{equation}
and
\begin{equation}\label{rho-n}
	{\rho_n}=(1-\rho_0)\big{[}(1-\frac{g_2}{g_1+g_2}\delta\beta\hbar\omega_0)e^{-\beta\hbar\omega_0}\big{]}^n
\end{equation}
up to the first order in $\delta \beta$. With straightforward algebra, the Lindblad model yields then the thermal conductance $G_{\rm th}^{\rm {Lindblad}}= P_1/\delta T$ for $\delta T \rightarrow 0$ accurately as
\begin{eqnarray}\label{G-th}
	G_{\rm th}^{\rm {Lindblad}}=\frac{g_1g_2}{g_1+g_2}\frac{\hbar^2\omega_0^3}{k_BT^2}n(\omega_0)[1+n(\omega_0)],
\end{eqnarray}
where $n(\omega_0)=1/(e^{\hbar\omega_0/(k_BT)}-1)$.

On the other hand, in the Landauer model shown in Fig.~\ref{fig1}~\,(b), two resistors $R_1$ and $R_2$ at temperatures $T_1$ and $T_2$, respectively, are coupled to each other via an element with impedance $Z(\omega)$. Fluctuating voltage $v_1$ induces a current $i_2$ governed by transconductance $G_{12}$ such that $i_2(\omega)=G_{12}(\omega)v_1(\omega)$. In a linear circuit, reciprocity yields $i_1(\omega)=G_{21}(\omega)v_2(\omega)$, where $ G_{21}(\omega)= G_{12}(\omega)$. The two noise sources are assumed to be uncorrelated, which then yields the net power between the two resistors as 
\begin{equation} \label{power_circ}
	P=\int_{0}^{\infty}\frac{d\omega}{2\pi} \big{[}S_{P_{12}}(\omega)-S_{P_{21}}(\omega)\big{]},
\end{equation}
where $S_{P_{ij}}(\omega)=R_j\,|G_{ij}(\omega)|^2\,S_{v_i}(\omega)$ with $i=1,\,2$. Here 
\begin{equation}\label{VNspec-1}
S_{v_i}(\omega) \simeq 4R_i \hbar\omega n_i(\omega)  
\end{equation}
is the voltage noise spectrum of resistor $i$ (excluding the zero point energy), with $n_i(\omega)=1/(e^{\hbar\omega/(k_BT_i)}-1)$ \cite{Schmidt}. Below we will again apply this result to the example circuit in Fig. \ref{fig2}. For this setup we find the transconductance
\begin{eqnarray} \label{G12_1}
	&&G_{12}(\omega)=G_{21}(\omega)=\\ &&\frac{Z_C}{(R_1+R_2+Z_1+Z_2)Z_C+R_1R_2+Z_1Z_2+Z_1R_2+Z_2R_1},\nonumber
\end{eqnarray}
where $Z_C=-i\tilde{Z}_{0}(\frac{\omega}{\tilde{\omega}_0}-\frac{\tilde{\omega}_0}{\omega})^{-1}$, with $\tilde{Z}_{0}=\sqrt{L/C}$, $\tilde{\omega}_0=1/\sqrt{LC}$, $Z_1=\frac{1}{i\omega C_1}$, and $Z_2=\frac{1}{i\omega C_2}$.
This presents a sharp peak of transmission $|G_{12}(\omega)|^2$ at $\omega=\tilde{\omega}_0$ in the weak coupling limit. Then we may approximate the integral of power in Eq. \eqref{power_circ} as
\begin{equation} \label{net power}
P=\int_{0}^{\infty}\frac{d\omega}{2\pi}\big{[}R_2|G_{12}(\omega)|^2S_{v_1}(\omega_0)-R_1|G_{12}(\omega)|^2S_{v_2}(\omega_0)\big{]},
\end{equation}
and it suffices to evaluate the integral of $|G_{12}(\omega)|^2$. In the leading order of the small coupling capacitances $C_1,C_2\ll C$ we have
\begin{equation} \label{G12-integral-1st}
	\int_{0}^{\infty}\frac{d\omega}{2\pi}|G_{12}(\omega)|^2=\frac{\tilde{\omega}_0^3 \tilde{Z}_{0}}{4\big{(}\frac{R_1}{C_1^2}+\frac{R_2}{C_2^2}\big{)}}.
\end{equation}
Then Eq.~\eqref{net power} yields
\begin{equation} \label{net power2}
P=\frac{R_1R_2\hbar\tilde{\omega}_0^4\tilde{Z}_0}{\frac{R_1}{C_1^2}+\frac{R_2}{C_2^2}}[n_1(\tilde{\omega}_0)-n_2(\tilde{\omega}_0)].
\end{equation}
We again consider the limit of a small temperature difference, $\delta T$, and make a linear approximation by setting $\beta_1 \equiv \beta $ and $\beta_2=\beta +\delta \beta$. In the weak-coupling limit, we further set $\omega_0=\tilde \omega_0$ and $Z_0=\tilde Z_0$. For the symmetric case, $C_1=C_2\equiv C_c$, we then obtain
\begin{equation} \label{net power3}
P=\frac{R_1R_2}{R_1+R_2}\frac{C_c^2}{Z_0C_\Sigma^2}\frac{\hbar^2{\omega}_0^3}{k_BT^2}n({\omega}_0)1+n({\omega}_0)]\delta T.
\end{equation}
Using the same definition of $g_j=\frac{R_j}{Z_0}(\frac{C_c}{C_\Sigma})^2$ as before, with $C_j$ replaced by $C_c$, we obtain $P=G_{\rm th}^{\rm {Circuit}}\delta T$, where we have
\begin{eqnarray}\label{G-th-circuit}
	G_{\rm th}^{\rm {Circuit}}=\frac{g_1g_2}{g_1+g_2}\frac{\hbar^2\omega_0^3}{k_BT^2}n(\omega_0)[1+n(\omega_0)],
\end{eqnarray}
which is exactly the same expression as Eq. \eqref{G-th} obtained above for the Lindblad model. This completes our demonstration that the two models coincide in the weak coupling limit.

\begin{figure}
	\centering
	\includegraphics [width=\columnwidth] {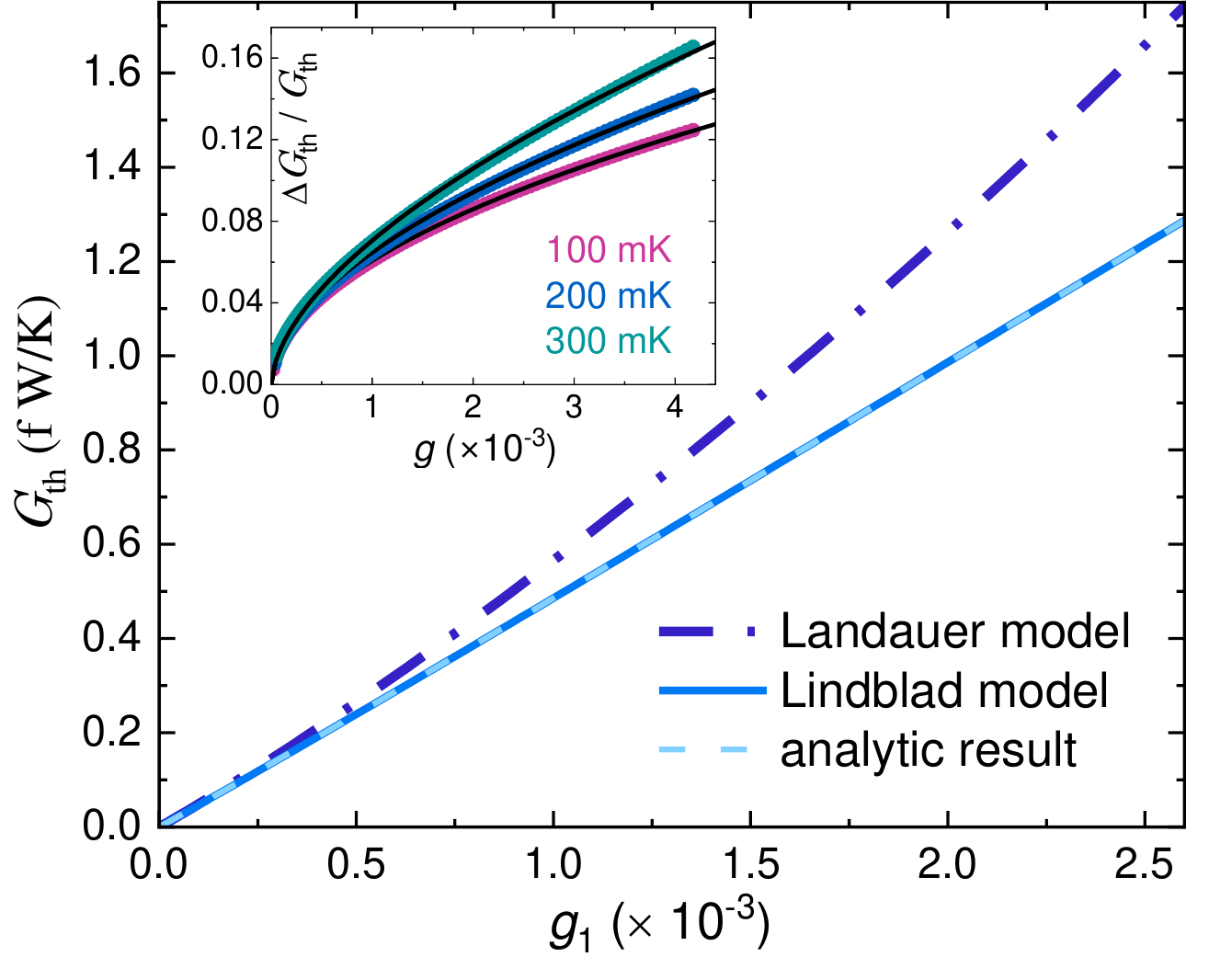}
	\caption{Thermal conductance in a circuit presented in Fig.~\ref{fig2} as a function of coupling $g_1$ for the two models along with the analytical result $P=G_{\rm th} \delta T$. The parameters are: $R_1=R_2=100\,\Omega$, $C=6\,C_1=2\,C_2=30\,$fF, and $L=4.2\,$nH. The inset plot shows $(G_{\rm th}^{\rm {Circuit}}-G_{\rm th}^{\rm {Lindblad}})/G_{\rm th}^{\rm {Circuit}}$ for three different temperatures as a function of coupling for the same circuit when the couplings are equal $C_1=\,C_2=5\,$fF. The black solid lines indicate fit to the numerical data with the function $P=ag^b$ where $b=0.50,\,0.54,\,$and $0.58$, for $100,\,200,\,$and $300\,$mK, respectively.
		\label{fig3}}
\end{figure}
{\it Numerical Implementation:} In this section, we describe the numerical procedure used to solve the governing equations and generate the results shown in Fig.~\ref{fig3}. We subsequently compare the steady-state powers obtained from the Lindblad model and circuit theory using direct numerical simulations, without relying on the approximations introduced above. In the Lindblad model, we calculate the transition rates from Eq.~\eqref{transition-rates}. Then, we solve the steady-state populations from Eq.~\eqref{densityM-n} setting $\dot{\rho}_n=0$ with normalization $\sum_{n=0}^{n_{\rm max}}{\rho}_n=1$, where $n_{\rm max}$ is chosen to be sufficiently large to include all the states with non-negligible population. Finally, the power is calculated using Eq.~\eqref{power} but again limiting the sum up to $n_{\rm max}$ levels. Circuit-theory predictions are obtained independently by numerically integrating Eq.~\eqref{net power}, using the thermal noise spectra of Eq.~\eqref{VNspec-1} and the transconductance defined in Eq.~\eqref{G12_1}.

In Fig.~\ref{fig3}, we present numerical results on comparing the power from the two models applied to the system in Fig.~\ref{fig2} without the approximations discussed earlier. The parameters in the modeling are taken to be representative for experiments on superconducting quantum circuits. We see that, indeed, the Lindblad model and the analytical approximation coincide over the whole range of the magnitude of coupling $g_1$. The ratio $g_2/g_1=3$ is maintained over the whole set of curves. At small values of coupling the two models coincide, as demonstrated by the analytic calculations above, whereas at larger values of coupling, the Lindblad weak coupling approximation falls below the exact circuit (Landauer) result. The inset of Fig.~\ref{fig3} demonstrates similar results, but now at three different temperatures. Moreover, we plot there the relative difference between the results from the circuit model and the weak coupling approximation as a function of $g\equiv g_1=g_2$. We see from the fits that the relative error of the Lindblad weak coupling model scales approximately as $\sqrt{g}$, a point that deserves further consideration.

Coincidence of the two discussed approaches for a linear circuit in the weak coupling limit served as a sanity check: at the end the two produced an identical analytic expression for thermal conductance of the circuit in Fig. \ref{fig2}. Yet, our work revealed that the requirement of weak coupling is quite restrictive if one wants to reach quantitative predictions by the Lindblad method. For our example circuit, a coupling constant of $g\sim 10^{-3}$ yields about 10\% underestimate for thermal conductance. This error grows approximately as $g^{1/2}$, which means that with $g\sim 0.01$, quite a common value in experiments on superconducting qubits, the error that the Lindblad approach yields is about 30\%. 

As discussed already earlier, the Lindblad approach is the method that can be applied to the study of anharmonic elements, i.e. non-linear circuits, which are in the focus of qubit-based architectures. As a concrete example, we can consider a quantum thermal diode, which produces non-reciprocal heat current in a configuration of Fig.~\ref{fig2}\,(a), if one replaces the harmonic oscillator by a qubit, ideally a two-level system. With the same procedure as described in Eqs.~\eqref{transition-rates}-\eqref{power}, but with states $n=0$, and $1$, we obtain the power to each bath $j$ as $P_j=\hbar\omega_Q(\rho_e\Gamma_\downarrow^{(j)}-\rho_g\Gamma_\uparrow^{(j)})$. Here, $\rho_e=\Gamma_\downarrow/(\Gamma_\downarrow+\Gamma_\uparrow)$, and $\rho_g=1-\rho_e$ are the population of the excited state and the ground state of the qubit, respectively, $\hbar\omega_Q$ is the energy-level spacing of the qubit and $\Gamma_{\downarrow\,(\uparrow)}=\Gamma_{\downarrow\,(\uparrow)}^{(1)}+\Gamma_{\downarrow\,(\uparrow)}^{(2)}$. Then the ratio of the forward and backward power is given by
\begin{equation}\label{rectification}
  \frac{P_1}{P_2}=\frac{g_1+g_2\coth(\frac{\beta_1\hbar\omega_Q}{2})\tanh(\frac{\beta_2\hbar\omega_Q}{2})}{g_1\coth(\frac{\beta_1\hbar\omega_Q}{2})\tanh(\frac{\beta_2\hbar\omega_Q}{2})+g_2}.  
\end{equation}
This demonstrates non-reciprocity for an asymmetric system when $g_1\neq g_2$~\cite{segal,segal2,Ojanen,Sothmann,Sanpera,Jsenior,Bibek,Giazotto}, which result is unattainable by the circuit model which forbids any diode-like behavior due to the reciprocity $G_{12}=G_{21}$ (see Eq.~\eqref{G12_1}). For a quantitative thermal analysis of these circuits beyond the very weak coupling regime, one then needs to go beyond the Lindblad master equation. 

\section{Acknowledgments} We gratefully acknowledge valuable discussions with Jukka Pekola. This work has received funding from the European Union’s Research and Innovation Programme, Horizon Europe, under the Marie Sklodowska-Curie Grant Agreement No. 101150440 (TcQTD)

\end{document}